\documentclass{article}
\usepackage{graphics}
\usepackage{epsfig}

\begin{document}

\begin{center} {\large \bf Polarization Properties of the "Photon Pistol"}
\end{center}
\bigskip
\begin{center}
\textbf{V.A. Reshetov}\\
\bigskip
\textit{Department of General and Theoretical Physics, Tolyatti
State University, 14 Belorousskaya Street, 446667 Tolyatti, Russia}
\end{center}

\begin{abstract}
The deterministic single-photon emission by means of STIRAP through
the atoms with degenerate levels is studied. The expression for the
polarization matrix of the emitted photon is obtained and its
dependence on the polarization of the driving laser field and on the
initial atomic state is examined.
\end{abstract}

\section{Introduction}

The application of quantum optical devices in quantum information
processing, specifically the efforts in constructing quantum
memories \cite{n1,n2,n3,n4} and quantum networks \cite{n5,n6,n7}, is
an active research area. The necessary part of such devices,
employing the photons as flying q-bits, is the deterministic
single-photon emitter - the "photon pistol". The most promising
scheme for the controlled generation of a single photon is based on
the technique of vacuum stimulated Raman scattering involving
adiabatic passage (STIRAP), proposed in \cite{n8} and then realized
experimentally in \cite{n9,n10,n11,n12}. In these experiments the
three-level $\Lambda$ - type atom was adiabatically passing through
a high-finesse cavity, one branch of the atomic $\Lambda$ - type
transitions was in resonance with the quantized cavity field, while
the other one was driven by the coherent laser field. In course of
interaction with the fields inside the cavity the atom emitted a
single photon. The two polarization degrees of freedom of the
emitted photon provide the most natural way to encode the q-bit,
however, in the early experiments \cite{n9,n10} the polarization of
the emitted photons was not controlled, nor detected, and the
degeneracy of resonant levels was not taken into account. In the
subsequent experiments \cite{n11,n12} the magnetic field directed
along the cavity axis was applied to lift the degeneracy of atomic
levels and the frequencies of the driving field and of the cavity
modes were adjusted to realize the non-degenerate $\Lambda$-scheme
with three Zeeman states, which allowed to produce the circularly
polarized photons. The objective of this paper is to calculate the
polarization state of the emitted photon and to study its dependence
on the polarization of the driving laser field and on the atomic
initial state for the arbitrary values of the level angular momenta
and to look for the schemes enabling to produce photons with
tailored polarizations.

The STIRAP with the three-level atom with non-degenerate levels is
well described in the reviews and textbooks (see, e.g.,
\cite{n13,n14}). In case of non-degenerate levels there exists the
only dark state -- the superposition of the ground state $a$ and
some metastable state $b$ -- uncoupled to the excited level $c$. In
course of Raman scattering the atom is adiabatically transferred
from the initial state $a$ to the target state $b$. The STIRAP with
degenerate atomic levels and with classical coherent resonant fields
was studied in \cite{n15}, were it was shown that unlike the
non-degenerate case the population from the initial ground state $a$
is not always totally transferred to the target level $b$, but only
if the order of degeneracy of this target state is not less, than
the order of degeneracy of the initial state. In our previous papers
\cite{n16,n17} we have studied the single-photon emission via
non-adiabatic vacuum stimulated Raman scattering of short pulses on
the atoms with degenerate levels and with totally or partially
resolved hyperfine structure, the polarization of the cavity mode
was assumed to be well-determined and the dependence of the photon
emission probability on the mutual orientation of polarizations of
the driving pulse and that of the cavity mode was calculated. In the
present paper the adiabatic Raman scattering (STIRAP) is considered,
and it is assumed that the cavity equally sustains both polarization
modes.

In section 2 the interaction model and the evolution operator for
this model in the adiabatic approximation are described, while in
sections 3 and 4 the instantaneous eigenvectors of the interaction
operator, which determine the evolution operator, are constructed
and classified. In case of degenerate levels not only the number of
these eigenvectors increases, but there appear the new types of
these eigenvectors, non-existing in case of non-degenerate levels,
like the dark states, which atomic part belongs to only one of the
lower levels $a$ or $b$, and the bright states, which couple the
excited level $c$ with only one of the lower levels $a$ or $b$. In
section 5 the formula for the polarization matrix of the emitted
photon is obtained and in section 6 this formula is used for
calculation of the photon polarization matrix for the transitions
with the angular momenta $J_{a}=J_{c}=3$, $J_{b}=2$ and
$J_{b}=J_{c}=1$, $J_{a}=2$, corresponding to the transitions between
the hyperfine structure components of the electronic levels
$5S_{1/2}$ and $5P_{3/2}$ of the $^{85}Rb$ and $^{87}Rb$ atoms,
which were employed in the experiments \cite{n9,n10,n11,n12,n18}.

\begin{figure}[t]\center
\includegraphics[width=7cm]{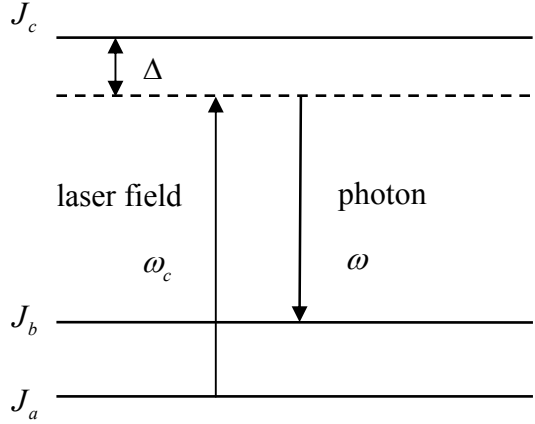}
\caption{The level diagram.}
\end{figure}

\section{Evolution operator}

Let us consider the coherent laser field with the carrier frequency
$\omega_{c}$, which is in resonance with the frequency $\omega_{c0}$
of an optically allowed transition $J_{a}\rightarrow J_{c}$ between
the ground state $J_{a}$ and the excited state $J_{c}$, while the
quantized cavity field with the carrier frequency $\omega$ is in
resonance with the frequency $\omega_{0}$ of an optically allowed
transition $J_{b}\rightarrow J_{c}$ between the long-lived state
$J_{b}$ and the same excited state $J_{c}$ (Fig.1). Here $J_{a}$,
$J_{b}$ and $J_{c}$ are the values of the angular momenta of the
levels. The coherent laser field is characterized by the electric
field strength
     \begin{equation}\label{q1}
\textbf{E}_{c}=e_{c}(t)\textbf{l}_{c} e^{-i\omega_{c}t}+ c.c.,
     \end{equation}
while the quantized field of the cavity in the interaction
representation is described by the operator:
     \begin{equation}\label{q2}
\hat{\textbf{E}}=e(t) (\hat{a}_{1}\textbf{l}_{1} +
\hat{a}_{2}\textbf{l}_{2}) e^{-i\omega t}+h.c.,
     \end{equation}
where $e_{c}(t)$ and $\textbf{l}_{c}$ are the slowly varying
amplitude and the unit polarization vector of the laser field,
$e(t)$ is the slowly varying amplitude of the cavity field,
$\textbf{l}_{1}$ and $\textbf{l}_{2}$ are the two unit orthogonal
vectors of the two polarization modes of this field, $\hat{a}_{1}$
and $\hat{a}_{2}$ are the photon annihilation operators for this
modes. The temporal dependence $e(t)$ of the cavity field amplitude
appears due to the motion of the atom through the cavity or to some
tailored alteration of cavity parameters.  The equation for the
slowly-varying density matrix $\hat{\rho}$ of the system, which
consists of a single three-level atom and two-mode cavity field, in
the rotating-wave approximation and in case of Raman resonance
$\omega_{c0}-\omega_{c}=\omega_{0}-\omega=\Delta$ is as follows:
     \begin{equation}\label{q3}
 \frac{d}{dt} \hat{\rho} =
 \frac{i}{2}\left[\hat{V}(t),\hat{\rho}\right],
     \end{equation}
     \begin{equation}\label{q4}
\hat{V}(t) = -2\Delta\hat{P}_{c} + \hat{G}(t) + \hat{G}^{\dag}(t),
     \end{equation}
     \begin{equation}\label{q5}
\hat{G}(t) = \Omega_{a}(t)\hat{g}_{a} + \Omega_{b}(t)\hat{g}_{b},
     \end{equation}
     \begin{equation}\label{q6}
\hat{g}_{b}=\hat{g}_{b1}\hat{a}_{1}^{\dag}+
\hat{g}_{b2}\hat{a}_{2}^{\dag}.
     \end{equation}
Here $\hat{P}_{c}$ is the projector on the subspace of the atomic
excited level $J_{c}$, $\Omega_{a}(t)=2|d_{a}|e_{c}(t)/\hbar$ and
$\Omega_{b}(t)=2|d_{b}|e(t)/\hbar$ are the reduced Rabi frequencies
for the coherent laser field and for the cavity field,
$d_{a}=d(J_{a}J_{c})$ and $d_{b}=d(J_{b}J_{c})$ are the reduced
matrix elements of the electric dipole moment operator for the
transitions $J_{a}\rightarrow J_{c}$ and $J_{b}\rightarrow J_{c}$,
while
    \begin{equation}\label{q7}
  \hat{g}_{a}=\hat{\textbf{g}}_{a}\textbf{l}_{c}^{*},~
  \hat{g}_{bi}=\hat{\textbf{g}}_{b}\textbf{l}^{*}_{i},~i=1,2,
    \end{equation}
$\hat{\textbf{g}}_{a}$ and $\hat{\textbf{g}}_{b}$ are the
dimensionless electric dipole moment operators for the transitions
$J_{a}\rightarrow J_{c}$ and $J_{b}\rightarrow J_{c}$. These
operators are expressed through Wigner 3J-symbols and partial atomic
operators
$$\hat{P}^{J_{\alpha}J_{\beta}}_{m_{\alpha}m_{\beta}} =
|J_{\alpha}m_{\alpha}><J_{\beta}m_{\beta}|,~\alpha,\beta = a,b,c, $$
in a following way \cite{n19}:
    \begin{equation}\label{q8}
  \hat{g}_{a}=\sum_{m_{a},m_{c},q}
(-1)^{J_{a}-m_{a}}l_{cq}^{*}\left(\matrix{J_{a}&1&J_{c}  \cr
-m_{a}&q&m_{c}}\right)\hat{P}^{J_{a}J_{c}}_{m_{a}m_{c}},
    \end{equation}
    \begin{equation}\label{q9}
  \hat{g}_{bi}=\sum_{m_{b},m_{c},q}
(-1)^{J_{b}-m_{b}}l_{iq}^{*}\left(\matrix{J_{b}&1&J_{c}  \cr
-m_{b}&q&m_{c}}\right)\hat{P}^{J_{b}J_{c}}_{m_{b}m_{c}},
    \end{equation}
where $l_{cq}$ and $l_{q}$ are the circular components of
polarization vectors $\textbf{l}_{c}$ and $\textbf{l}_{i}$.

The solution of the equation (\ref{q3}) is expressed through the
evolution operator $\hat{S}(t)$:
    \begin{equation}\label{q10}
   \hat{\rho}(t)=\hat{S}(t)\hat{\rho}(0)\hat{S}^{+}(t).
    \end{equation}
In the adiabatic approximation (see, e.g., \cite{n13}) the evolution
operator $\hat{S}(t)$ is defined by the instantaneous eigenvectors
$|v_{k}(t)>$ and eigenvalues $\lambda_{k}(t)$ of the interaction
operator $\hat{V}(t)$:
    \begin{equation}\label{q13}
\hat{S}(t)=\sum_{k}\exp\{i\phi_{k}(t)\} |v_{k}(t)><v_{k}(0)|,
    \end{equation}
    \begin{equation}\label{q14}
\phi_{k}(t)=\frac{1}{2}\int_{0}^{t}\lambda_{k}(t')dt'.
    \end{equation}

\section{Bright states}

Since only the processes of a single photon emission are discussed
in the present paper it is sufficient to limit the system space to
the subspace with the basis vectors $|J_{a}m_{a}>|0,0>$,
$|J_{b}m_{b}>|1,0>$, $|J_{b}m_{b}>|0,1>$, $|J_{c}m_{c}>|0,0>$, where
$|J_{a}m_{a}>$, $|J_{b}m_{b}>$ and $|J_{c}m_{c}>$ denote the atomic
Zeeman states, while $|n_{1},n_{2}>$ ($n_{1,2}=0,1$) are the field
number states with $n_{1}$ photons in the first polarization mode
and $n_{2}$ -- in the second. This subspace with the dimension
$N=2(J_{a}+2J_{b}+J_{c}+2)$ constitutes the invariant subspace of
the interaction operator $\hat{V}(t)$, so that its matrix represents
itself a square hermitian $N\times N$ matrix. The states $|a>|0,0>$
and $|c>|0,0>$, which atomic part belongs to the level $a$ or $c$,
may be represented as columns with $2J_{a}+1$ or $2J_{c}+1$ elements
correspondingly, while the states $|b_{1}>|1,0>+|b_{2}>|0,1>$, which
atomic part belongs to the level $b$, may be represented as columns
with $2(2J_{b}+1)$ elements:
$$|b_{1}>|1,0>+|b_{2}>|0,1> = \left(\matrix{|b_{1}> \cr
|b_{2}>}\right).$$ Then the operator $\hat{g}_{a}$ will be
represented by the $(2J_{a}+1)\times (2J_{c}+1)$ matrix, while the
operator $\hat{g}_{b}$ will be represented by the $2(2J_{b}+1)\times
(2J_{c}+1)$ matrix
$$\hat{g}_{b} = \left( \matrix{
\hat{g}_{b1} \cr \hat{g}_{b2}} \right),$$ were each block
$\hat{g}_{bi}$ ($i=1,2$) represents itself a $(2J_{b}+1)\times
(2J_{c}+1)$ matrix.

In order to find out the instantaneous eigenvectors $|v_{k}(t)>$ and
eigenvalues $\lambda_{k}(t)$ of the interaction operator
$\hat{V}(t)$ let us start with the eigenvectors of the operator
$\hat{G}^{\dag}(t)\hat{G}(t)$, which acts at the subspace of the
upper atomic level $c$.  Let us denote as $|D_{n}^{c}>$ the states,
if there are any, which are simultaneously the eigenvectors of both
operators $\hat{g}_{a}^{\dag}\hat{g}_{a}$ and
$\hat{g}_{b}^{\dag}\hat{g}_{b}$ with zero eigenvalues:
    \begin{equation}\label{q15}
\hat{g}_{a}^{\dag}\hat{g}_{a}|D_{n}^{c}> =
\hat{g}_{b}^{\dag}\hat{g}_{b}|D_{n}^{c}> = 0,
     \end{equation}
the number of such sates being $N_{c}^{d}$. These states remain
uncoupled to the lower atomic levels $a$ and $b$. Next, let us
consider the states $|C_{n}^{a}>$, which are the eigenvectors of
operator $\hat{g}_{a}^{\dag}\hat{g}_{a}$ with non-zero eigenvalues
and at the same time the eigenvectors of the operator
$\hat{g}_{b}^{\dag}\hat{g}_{b}$ with zero eigenvalues:
    \begin{equation}\label{q16}
\hat{g}_{a}^{\dag}\hat{g}_{a}|C_{n}^{a}> =c_{an}^{2} |C_{n}^{a}>,~
c_{an}>0,~ \hat{g}_{b}^{\dag}\hat{g}_{b}|C_{n}^{a}> = 0,
     \end{equation}
the number of such sates being $N_{c}^{a}$. These states are coupled
to the lower atomic level $a$ only. Similarly, the states
$|C_{n}^{b}>$, coupled to the lower atomic level $b$ only, are the
eigenvectors of the operator $\hat{g}_{b}^{\dag}\hat{g}_{b}$ with
non-zero eigenvalues, which are at the same time the eigenvectors of
the operator $\hat{g}_{a}^{\dag}\hat{g}_{a}$ with zero eigenvalues:
    \begin{equation}\label{q17}
\hat{g}_{b}^{\dag}\hat{g}_{b}|C_{n}^{b}> =c_{bn}^{2} |C_{n}^{b}>,~
c_{bn}>0,~ \hat{g}_{a}^{\dag}\hat{g}_{a}|C_{n}^{b}> = 0,
     \end{equation}
the number of such sates being $N_{c}^{b}$. Finally, let us consider
the states $|C_{n}(t)>$, which satisfy the inequations:
$$\hat{g}_{a}|C_{n}(t)> \neq 0,~ \hat{g}_{b}|C_{n}(t)> \neq 0,$$
at any time. These states, coupled to both lower atomic levels $a$
and $b$, may be obtained as the eigenvectors of the operator
$\hat{G}^{\dag}(t)\hat{G}(t)$ with non-zero eigenvalues at any time:
    \begin{equation}\label{q18}
\hat{G}^{\dag}(t)\hat{G}(t)|C_{n}(t)>=c_{n}^{2}(t)|C_{n}(t)>,~
c_{n}(t)>0,
    \end{equation}
the number of such sates being $N_{c}$. All the temporally
independent $N_{d}^{c}$ states $|D_{n}^{c}>$, $N_{c}^{a,b}$ states
$|C_{n}^{a,b}>$ and temporally dependent $N_{c}$ states $|C_{n}(t)>$
constitute the complete
($N_{d}^{c}+N_{c}^{a}+N_{c}^{b}+N_{c}=2J_{c}+1$) orthonormal set of
states with the atomic part belonging to the atomic level $c$ at any
instant of time.

The bright states
    \begin{equation}\label{q19}
|F_{n}^{a,b}> =
\frac{1}{c_{a,bn}}\hat{g}_{a,b}|C_{n}^{a,b}>,~n=1,...,N_{c}^{a,b},
    \end{equation}
    \begin{equation}\label{q20}
|F_{n}(t)> = \frac{1}{c_{n}(t)}\hat{G}(t)|C_{n}(t)>,~n=1,...,N_{c},
    \end{equation}
coupled to the states $|C_{n}^{a,b}>$ and $|C_{n}(t)>$ by electric
dipole transitions, also form an orthonormal set of states, as it
follows from (\ref{q19})-(\ref{q20}). The atomic part of the
temporally independent states $|F_{n}^{a,b}>$ belongs to the
subspace of the only one lower level $a$ or $b$, while the atomic
part of the temporally dependent states $|F_{n}(t)>$ belongs to the
subspace of both lower levels $a$ and $b$.

With the introduction of the states $|D_{n}^{c}>$, $|C_{n}^{a,b}>$,
$|C_{n}(t)>$ and $|F_{n}^{a,b}>$, $|F_{n}(t)>$, the interaction
operator $\hat{V}(t)$ may be easily diagonalized. It has $N_{c}^{d}$
eigenvectors $|D_{n}^{c}>$, then $2(N_{c}^{a}+N_{c}^{b})$
eigenvectors $|V_{a,bn}^{(\pm)}(t)>$, which are linear
superpositions of states $|F_{n}^{a,b}>$ and $|C_{n}^{a,b}>$, and
$2N_{c}$ eigenvectors $|V_{n}^{(\pm)}(t)>$ which are linear
superpositions of states $|F_{n}(t)>$ and $|C_{n}(t)>$. All these
states $|D_{n}^{c}>$, $|V_{a,bn}^{(\pm)}(t)>$ and
$|V_{n}^{(\pm)}(t)>$ constitute the orthonormal set of
$N^{f}=2(2J_{c}+1)-N^{d}_{c}$ eigenvectors of the operator
$\hat{V}(t)$ with non-zero eigenvalues.

\section{Dark states}

The other $N^{d}=N-N^{f}$ eigenvectors $|D_{k}(t)>$ of the
interaction operator $\hat{V}(t)$  obtain zero eigenvalues:
   $$\hat{V}(t)|D_{k}(t)> = 0,~ k=1,...,N_{d}.$$
These states with the atomic part belonging to the subspace of the
lower atomic levels $a$ and $b$ remain uncoupled to the upper level
$c$. These states -- dark states -- satisfy the equation:
    \begin{equation}\label{q21}
\hat{G}^{\dag}(t)|D_{k}(t)> = 0,~ k=1,...,N_{d}.
    \end{equation}
As it follows from (\ref{q21}), all the dark states $|D_{k}(t)>$ are
orthogonal to all the states $|D_{n}^{c}>$, $|V_{a,bn}^{(\pm)}(t)>$
and $|V_{n}^{(\pm)}(t)>$.

Among all the dark states let us distinguish first the dark states
$|D_{k}^{a}>$, which atomic part belongs to the lower level $a$
only. These states are time independent and may be obtained as the
eigenvectors of the operator $\hat{g}_{a}\hat{g}_{a}^{\dag}$ with
zero eigenvalues:
     \begin{equation}\label{q22}
\hat{g}_{a}\hat{g}_{a}^{\dag}|D_{k}^{a}> = 0,~ k=1,...,N_{a}^{d}.
     \end{equation}
All the other eigenvectors $|A_{n}>$ of the operator
$\hat{g}_{a}\hat{g}_{a}^{\dag}$ with non-zero eigenvalues, which
satisfy the equation
     \begin{equation}\label{q23}
\hat{g}_{a}\hat{g}_{a}^{\dag}|A_{n}> = a_{n}^{2}|A_{n}>,~ a_{n}>0,~
n=1,...,N_{a},
     \end{equation}
constitute the subspace orthogonal to the subspace of the dark
states $|D_{k}^{a}>$, its dimension being $N_{a}=
2J_{a}+1-N_{a}^{d}$. Similarly the dark states $|D_{k}^{b}>$, which
atomic part belongs to the lower level $b$ only, may be obtained as
the eigenvectors of the operator $\hat{g}_{b}\hat{g}_{b}^{\dag}$
with zero eigenvalues:
     \begin{equation}\label{q24}
\hat{g}_{b}\hat{g}_{b}^{\dag}|D_{k}^{b}>  = 0,~ k=1,...,N_{b}^{d}.
     \end{equation}
The $N_{b}= 2(2J_{b}+1)-N_{b}^{d}$ eigenvectors $|B_{n}>$ of the
operator $\hat{g}_{b}\hat{g}_{b}^{\dag}$ with non-zero eigenvalues:
     \begin{equation}\label{q25}
\hat{g}_{b}\hat{g}_{b}^{\dag}|B_{n}> = b_{n}^{2}|B_{n}>,~ b_{n}>0,~
n=1,...,N_{b},
     \end{equation}
constitute the subspace orthogonal to the subspace of the dark
states $|D_{k}^{b}>$.

Let us now consider the dark states $|D_{k}^{ab}(t)>$, which atomic
part belongs to both atomic lower levels $a$ and $b$. These states
satisfy the equation
    \begin{equation}\label{q26}
\hat{G}^{\dag}(t)|D^{ab}_{k}(t)> = 0,~ k=1,...,N_{ab}^{d},
    \end{equation}
while
$$\hat{g}_{a}^{\dag}|D_{k}^{ab}(t)> \neq 0,~
\hat{g}_{b}^{\dag}|D_{k}^{ab}(t)> \neq 0.$$ The temporal dependence
of these states may be immediately obtained from the equation
(\ref{q26}):
 $$|D^{ab}_{k}(t)> = Z_{k}(t)[\Omega_{a} (t) |B_{k}^{d}> - \Omega_{b} (t)
|A_{k}^{d}>],$$ where $Z_{k}(t)$ is the normalization factor, while
$|A_{k}^{d}>$ and $|B_{k}^{d}>$ are temporally independent states,
which atomic parts belong to the levels $a$ and $b$ correspondingly
and which satisfy the equation
    \begin{equation}\label{q27}
\hat{g}_{a}^{\dag}|A_{k}^{d}> = \hat{g}_{b}^{\dag}|B_{k}^{d}> \neq
0.
    \end{equation}
Introducing the matrix
    \begin{equation}\label{q28}
\hat{D}_{b} = \sum_{n=1}^{N_{b}}\frac{1}{b_{n}^{2}} |B_{n}><B_{n}|,
    \end{equation}
containing only the eigenvectors $|B_{n}>$ of matrix
$\hat{g}_{b}\hat{g}_{b}^{\dag}$ with non-zero eigenvalues, we may
write the equation (\ref{q27}) as follows:
    \begin{equation}\label{q29}
|B_{k}^{d}> = \hat{D}_{ba}|A_{k}^{d}>,~ \hat{D}_{ba} =
\hat{D}_{b}\hat{g}_{b}\hat{g}_{a}^{\dag}.
    \end{equation}
Now we may define the orthonormal set of states $|A_{k}^{d}>$
($k=1,...,N_{ab}^{d}$) as the eigenvectors of the hermitian matrix
$\hat{D}_{ba}^{\dag}\hat{D}_{ba}$ with non-zero eigenvalues:
    \begin{equation}\label{q30}
\hat{D}_{ba}^{\dag}\hat{D}_{ba}|A_{k}^{d}> = a_{dk}^{2}|A_{k}^{d}>,~
a_{dk}>0.
    \end{equation}
Then, as it follows from (\ref{q29}) and (\ref{q30}), the states
$a_{dk}^{-1}|B_{k}^{d}>$ also constitute the orthonormal set, so
that the orthonormal set of dark states $|D^{ab}_{k}(t)>$ may be
expressed through the eigenvectors $|A_{k}^{d}>$ of the equation
(\ref{q30}) by the formula:
    \begin{equation}\label{q31}
|D^{ab}_{k}(t)> = \frac{\Omega_{a}(t)\hat{D}_{ba}|A_{k}^{d}> -
\Omega_{b}(t)|A_{k}^{d}>}
{\sqrt{\Omega_{b}^{2}(t)+a_{dk}^{2}\Omega_{a}^{2}(t)}}.
    \end{equation}

\section{Single photon emission}

All the eigenvectors of the interaction operator $\hat{V}(t)$,
comprising the set of states $|D_{n}^{c}>$, $|V_{a,bn}^{(\pm)}(t)>$,
$|V_{n}^{(\pm)}(t)>$ and the set of dark states $|D_{k}^{a,b}>$,
$|D_{k}^{ab}(t)>$, constitute the complete orthonormal set of
states, which determines the evolution operator (\ref{q13}) in the
adiabatic approximation. Initially the atom is at its lower level
$a$, while the cavity field is in its vacuum state, the initial
atomic density matrix of the atom+field system being
$\hat{\rho}_{0}^{a}$. The classical coherent laser field is
adiabatically switched on, while the interaction with the quantum
field is adiabatically switched off in the time interval $T$, so
that:
$$\Omega_{a}(0)=0,~ \Omega_{a}(T)=\Omega_{a0},$$
$$\Omega_{b}(0)=\Omega_{b0},~ \Omega_{b}(T)=0.$$
The atomic part of the initial states $|V_{n}^{(\pm)}(0)>$ belongs
to the atomic levels $b$ and $c$ only, so that only the states
$|V_{an}^{(\pm)}(t)>$ and the dark states $|D_{k}^{a}>$ and
$|D_{k}^{ab}(t)>$ will contribute to the evolution operator
(\ref{q13}). After the STIRAP process is finished at $t=T$ the
atomic part of the states $|V_{an}^{(\pm)}(T)>$ belongs to the
levels $a$ and $c$, the atomic part of the states $|D_{k}^{a}>$
belongs to the level $a$ and the atomic part of the states
$|D_{k}^{ab}(T)>$ belongs to the level $b$, so that only the
presence of the dark states $|D_{k}^{ab}(t)>$ in the evolution
operator contributes to the probability of the photon emission,
while the presence of the states $|V_{an}^{(\pm)}(t)>$ and
$|D_{k}^{a}>$ in the evolution operator reduces this probability.
The total probability $w$ of the single photon emission is
    \begin{equation}\label{q32}
w = <1,0|\hat{\rho}_{F}|1,0> + <0,1|\hat{\rho}_{F}|0,1>,
    \end{equation}
while the polarization of the photon is described by the $2\times 2$
polarization matrix
    \begin{equation}\label{q33}
\sigma=\frac{1}{w}\left(\matrix{<1,0|\hat{\rho}_{F}|1,0> &
<1,0|\hat{\rho}_{F}|0,1> \cr <0,1|\hat{\rho}_{F}|1,0> &
<0,1|\hat{\rho}_{F}|0,1>} \right).
    \end{equation}
Here
\begin{equation}\label{q34}
\hat{\rho}_{F} = tr_{A}\left\{\hat{U}\hat{\rho}_{0}^{a}
\hat{U}^{\dag}\right\},~ \hat{U} = \hat{D}_{ba}\hat{P}_{a}^{d},
    \end{equation}
the trace is carried out over the atomic variables, the operator
$\hat{D}_{ba}$ is defined by (\ref{q29}) and
    \begin{equation}\label{q35}
\hat{P}_{a}^{d} = \sum_{k=1}^{N_{ab}^{d}} \frac{1}{a_{dk}}
|A_{k}^{d}><A_{k}^{d}|,
    \end{equation}
where $|A_{k}^{d}>$ and $a_{dk}$ are defined by (\ref{q30}).

\section{Discussion}

Let the quantization axis $Z$ be directed along the cavity axis and
let us choose the two unit polarization vectors of the quantum field
as the two unit vectors along the Cartesian axes $X$ and $Y$:
$$\textbf{l}_{1}=\textbf{e}_{x},~\textbf{l}_{2}=\textbf{e}_{y},$$
while the driving field propagates along the axis $Y$ so that its
polarization vector $\textbf{l}_{c}$ belongs to the plane $XZ$. In
case of linearly polarized driving field
$$\textbf{l}_{c}=\textbf{e}_{z}\cos\psi+\textbf{e}_{x}\sin\psi.$$
The photon polarization matrix (\ref{q33}) may be expressed through
the Stokes parameters $\xi_{n}$ ($n=1,2,3$) \cite{n20}:
$$\sigma = \frac{1}{2} \left( \matrix{1+\xi_{3}
 &\xi_{1}-i\xi_{2} \cr \xi_{1}
 +i\xi_{2}&1-\xi_{3}}\right),$$
which determine the photon total degree of polarization
$$P=\sqrt{\xi_{1}^{2}+\xi_{2}^{2}+\xi_{3}^{2}}.$$
In the experiments \cite{n9,n10} the resonant levels with angular
momenta $J_{a}=J_{c}=3$, $J_{b}=2$ were the hyperfine structure
components of the electronic levels $5S_{1/2}$ and $5P_{3/2}$ of the
$^{85}Rb$ atom, however the polarization of the driving field and
the initial atomic state were not specified. So let the atom be
initially in the equilibrium state
$$\hat{\rho}_{0}^{a}=\frac{\hat{P}_{a}}{2J_{a}+1}|0,0><0,0|,$$
where $\hat{P}_{a}$ is the projector on the subspace of the atomic
ground state $a$. Then we obtain from the numeric calculations based
on the formulae (\ref{q32})-(\ref{q35}), that the photon emission
probability $w=0.857$ is less than unity and does not depend on the
angle $\psi$ between the cavity axis $Z$ and the polarization vector
$\textbf{l}_{c}$ of the driving field, while the emitted photon is
almost unpolarized, its degree of polarization is zero $P=0$ with
the $\pi$-polarized driving field ($\psi=0$) and $P=0.07$ with the
$\sigma$-polarized field ($\psi=\pi/2$). Now let us assume that the
atom is initially prepared in a pure Zeeman state
$|J_{a}=3,m_{a}=0>$ with zero value of the angular momentum
projection $m_{a}$ on the quantization axis $Z$, like in the
experiment \cite{n18}. Then we obtain from (\ref{q32})-(\ref{q35})
the dependence of the photon emission probability $w$ on the angle
$\psi$ presented at Figure 2. In this case with the $\pi$-polarized
driving field ($\psi=0$) the photon is not emitted at all $w=0$,
while it is emitted with the unit probability $w=1$ with the
$\sigma$-polarized field ($\psi=\pi/2$) and at an angle $\psi=0.685$
rad ($39^{o}$), at $\psi=\pi/2$ its degree of polarization being
$P=0.43$ and at $\psi=0.685$ it is $P=0.80$.

In the experiments \cite{n11,n12} the magnetic field directed along
the cavity axis $Z$ was applied to lift the degeneracy of atomic
levels and the cavity field was tuned in resonance with the
transition between the Zeeman state $|F'=1,m_{F'}=0>$ of the
hyperfine structure component $F'=1$ of the excited $5P_{3/2}$
electronic level of $^{87}Rb$ atom and one of the Zeeman states
$|F=1,m_{F}=\pm 1>$ of the hyperfine structure component $F=1$ of
the ground level $5S_{1/2}$, while the linearly $\sigma$-polarized
driving field was in resonance with the transition between the same
Zeeman state $|F'=1,m_{F'}=0>$ of the excited level and the other
Zeeman state $|F=1,m_{F}=\mp 1>$ of the ground level, so that the
emitted photons were circularly polarized, as it is evident from
selection rules.

\begin{figure}[t]\center
\includegraphics[width=7cm]{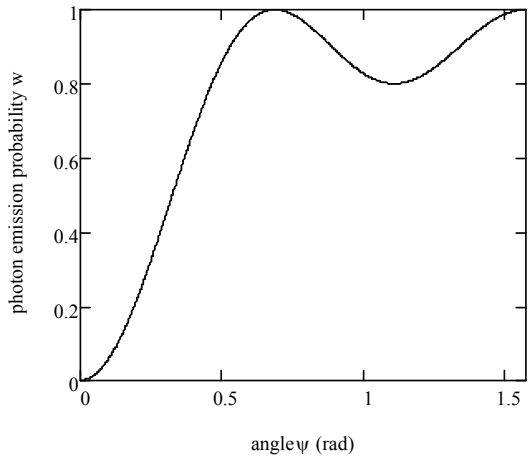}
\caption{The photon emission probability $w$ on the transitions with
the angular momenta $J_{a}=J_{c}=3$, $J_{b}=2$ versus the angle
$\psi$ between the cavity axis $Z$ and the polarization vector
$\textbf{l}_{c}$ of the driving field in case of initial pure atomic
state $m_{a}=0$ with zero angular momentum projection on the
quantization axis $Z$.}
\end{figure}
\begin{figure}[t]\center
\includegraphics[width=7cm]{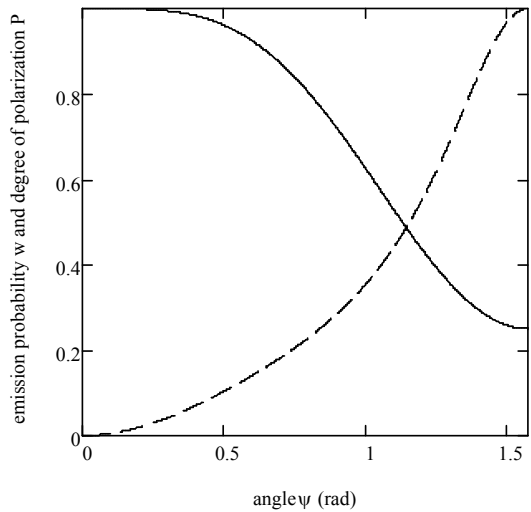}
\caption{The photon emission probability $w$ (solid line) and the
degree of polarization $P$ (dashed line) on the transitions with the
angular momenta $J_{b}=J_{c}=1$, $J_{a}=2$ versus the angle $\psi$
between the cavity axis $Z$ and the polarization vector
$\textbf{l}_{c}$ of the driving field in case of initial pure atomic
state $m_{a}=0$ with zero angular momentum projection on the
quantization axis $Z$.}
\end{figure}

In the experiment \cite{n18} the resonant levels with angular
momenta $J_{b}=J_{c}=1$, $J_{a}=2$ were the hyperfine structure
components of the electronic levels $5S_{1/2}$ and $5P_{3/2}$ of the
$^{87}Rb$ atom and the atom was initially prepared in the pure
Zeeman state $|J_{a}=2,m_{a}=0>$, while the driving field was
linearly $\pi$-polarized. The results of numeric calculations of the
photon emission probability $w$ and the degree of polarization $P$
versus the angle $\psi$ between the cavity axis $Z$ and the
polarization vector $\textbf{l}_{c}$ of the driving field for these
transitions are presented at Figure 3. In case of $\pi$-polarized
($\psi=0$) driving field, as it was in the experiment, the photon is
emitted with unit probability $w=1$, but it is fully unpolarized
$P=0$. The photon becomes unpolarized after the averaging over the
states of the atom, however in the experiment \cite{n18} the same
atom was driven twice in the time interval less than decoherence
time to produce the sequence of two photons with strongly correlated
polarizations. As it may be seen from Figure 3, with the same
experimental setup but with the driving field linearly
$\sigma$-polarized ($\psi=\pi/2$) the emitted photon will be fully
polarized $P=1$ (linearly polarized along the axis $Y$ as follows
from calculations), though the emission probability in this case is
reduced to $w=0.25$. To obtain the linearly polarized photons with
unit probability the scheme with lower values of angular momenta
$J_{a}=0$, $J_{b}=J_{c}=1$, will be suitable. In this case the
$\sigma$-polarized driving field will produce the photons linearly
polarized along the axis $Y$ with unit probability.

\section{Conclusions}

In the present paper the general formulae for calculation of the
polarization matrix of a single photon emitted in the microcavity by
a single three-level $\Lambda$-type atom with degenerate levels,
driven adiabatically by the classical laser field from the ground
state $a$ to the long-lived state $b$ through the excited state $c$,
are obtained for the arbitrary polarization of the driving field and
arbitrary values of the angular momenta $J_{a}$, $J_{b}$ and $J_{c}$
of the resonant atomic levels.

For the transitions with the angular momenta $J_{a}=J_{c}=3$,
$J_{b}=2$ and $J_{b}=J_{c}=1$, $J_{a}=2$, corresponding to the
transitions between the hyperfine structure components of the
electronic levels $5S_{1/2}$ and $5P_{3/2}$ of the $^{85}Rb$ and
$^{87}Rb$ atoms, which were employed in the experiments
\cite{n9,n10,n18}, and for the atom initially prepared at pure
Zeeman state with zero projection $m_{a}=0$ on the quantization
axis, the dependencies of the photon emission probability $w$ and
degree of polarization $P$ on the angle $\psi$ between the cavity
axis and the polarization vector $\textbf{l}_{c}$ of the driving
field were calculated numerically. In case of transitions with
$J_{a}=J_{c}=3$, $J_{b}=2$, the unit emission probability may be
obtained at the angles $\psi=39^{o}$ and $\psi=90^{o}$ with the
degrees of polarization $P=0.80$ and $P=0.43$ correspondingly. In
case of transitions with $J_{b}=J_{c}=1$, $J_{a}=2$, at $\psi=0$ the
fully unpolarized photon is emitted with unit probability, while at
$\psi=90^{o}$ fully linearly polarized photon is emitted with the
probability $w=0.25$.

The fully linearly polarized photon emitted with unit probability
may be obtained by means of $\sigma$-polarized driving field
($\psi=90^{o}$) with lower values of the level angular momenta
$J_{b}=J_{c}=1$, $J_{a}=0$.

{\bf Acknowledgements}

Author is indebted for financial support of this work to Russian
Ministry of Science and Education (grant 2407).


\begin{thebibliography}{00}

\bibitem{n1} A.I. Lvovsky, B.C. Sanders, W. Tittel, {\it Nature
Photonics} {\bf 3} (2009), 706-714.

\bibitem{n2} H.P. Specht, C. Nolleke, A. Reiserer, M.
Uphoff, E. Figueroa, S.Ritter, G. Rempe, {\it Nature} {\bf 473}
(2011), 190-193.

\bibitem{n3} M. Himsworth, P. Nisbet, J. Dilley,
G. Langfahl-Klabes, A. Kuhn, {\it Appl.Phys.B} {\bf 103} (2011),
579-589.

\bibitem{n4} B. Lauritzen, N. Timoney, N. Gisin, M. Afzelius, H. Riedmatten,
Y. Sun, R. M. Macfarlane, R. L. Cone, {\it Phys. Rev. B} {\bf 85}
(2012), 115111.

\bibitem{n5} H.J. Kimble, {\it Nature} {\bf 453} (2008),
1023-1030.

\bibitem{n6} C. Simon, H. Riedmatten, N. Gisin, {\it Rev.Mod.Phys.}
{\bf 83} (2011), 33-80.

\bibitem{n7} S. Riedl, M. Lettner, C. Vo, S. Baur, G. Rempe and S.
Durr, {\it Phys.Rev.A} {\bf 85} (2012), 022318.

\bibitem{n8} A.S. Parkins, P. Marte, P. Zoller and H.J. Kimble,
{\it Phys.Rev.Lett.} {\bf 71} (1993), 3095-3098.

\bibitem{n9} M. Hennrich, T. Legero, A. Kuhn and G. Rempe,
{\it Phys.Rev.Lett.} {\bf 85} (2000), 4872-4875.

\bibitem{n10} A. Kuhn, M. Hennrich and G. Rempe,
{\it Phys.Rev.Lett.} {\bf 89} (2002), 067901.

\bibitem{n11} T. Wilk, S.C. Webster, H.P. Specht, G. Rempe and  A. Kuhn,
{\it Phys.Rev.Lett.} {\bf 98} (2007), 063601.

\bibitem{n12} T. Wilk, H.P. Specht, S.C. Webster, G. Rempe, A. Kuhn,
{\it  J.Mod.Opt.} {\bf 54} (2007), 1569-1580.

\bibitem{n13} M. Fleischhauer, A. Imamoglu, J.P. Marangos,
{\it Rev.Mod.Phys.} {\bf 77} (2005), 633-673.

\bibitem{n14} M.O. Scully, M.S. Zubairy, Quantum Optics,
Cambridge University Press, Cambridge (1997).

\bibitem{n15} Z. Kis, A. Karpati, B.W. Shore and N.V. Vitanov,
{\it Phys.Rev.A}, {\bf 70} (2004), 053405.

\bibitem{n16} V.A. Reshetov and I.V. Yevseyev, {\it Laser Phys.Lett.}
{\bf 5} (2008), 608-613.

\bibitem{n17} V.A. Reshetov and I.V. Yevseyev, {\it Opt.Commun.}
{\bf 283} (2010), 2557-2560.

\bibitem{n18} T. Wilk, S.C. Webster, A. Kuhn, G. Rempe,
{\it Science} {\bf 317} (2007), 488-490.

\bibitem{n19} I.I. Sobelman, Introduction to the Theory of Atomic
Spectra, Pergamon, New York (1972).

\bibitem{n20} L.D. Landau, E.M. Lifshits, The classical theory of
fields, Pergamon, New York (1987).


\end{thebibliography}
\end{document}